\documentclass[twocolumn,superscriptaddress,aps,pra,10pt]{revtex4-2}

\usepackage{amsmath}
\usepackage[latin9]{inputenc}
\pdfoutput=1
\usepackage{hyperref}
\usepackage{orcidlink}
\usepackage{soul}
\usepackage{appendix}
\usepackage{graphicx}
\usepackage{amssymb}
\usepackage{gensymb}
\usepackage{subcaption}
\usepackage{ragged2e}

\usepackage{comment}

\usepackage[english]{babel}
\usepackage[T1]{fontenc}

\usepackage{silence}
\usepackage{nicematrix}
\usepackage{mathtools}
\usepackage{float}
\usepackage{graphicx}

\usepackage{comment}
\usepackage{bbold}

\definecolor{cobalt}{rgb}{0.0, 0.28, 0.67}

\def\id{{\mathbb I}}
\def\qed{\ensuremath{\hfill \blacksquare}}

\DeclareMathOperator\Tr{Tr}
\definecolor{azure}{rgb}{0.0, 0.5, 1.0}
\definecolor{cobalt}{rgb}{0.0, 0.28, 0.67}

\newtheorem{theo}{Theorem}

\newtheorem{lemma}[theo]{Lemma}

\makeatletter
\def\blfootnote{\xdef\@thefnmark{}\@footnotetext}
\makeatother

\begin{document}

\title{Certifying asymmetry in the configuration of three qubits}

\author{Abdelmalek Taoutioui
\orcidlink{0000-0002-4943-5529}}
\affiliation{HUN-REN Institute for Nuclear Research, P.O. Box 51, H-4001 Debrecen, Hungary}

\author{G\'abor Dr\'otos
\orcidlink{0000-0002-0900-5188}}
\affiliation{HUN-REN Institute for Nuclear Research, P.O. Box 51, H-4001 Debrecen, Hungary}
\affiliation{Instituto de F\'isica Interdisciplinar y Sistemas Complejos (CSIC-UIB), Campus UIB, Carretera de Valldemossa km 7,5, E-07122 Palma de Mallorca, Spain}

\author{Tam\'as V\'ertesi
\orcidlink{0000-0003-4437-9414}}
\affiliation{HUN-REN Institute for Nuclear Research, P.O. Box 51, H-4001 Debrecen, Hungary}

\begin{abstract}
Symmetry restrictions limit the types of tasks that can be achieved with a given set of quantum states. Therefore, any breaking of these symmetries could potentially be exploited as a resource for quantum communication. Here we demonstrate this operationally by certifying asymmetry in the configuration of the Bloch vectors of a set of three unknown qubit states within the dimensionally bounded prepare-and-measure scenario. To do this, we construct a linear witness from three simpler witnesses as building blocks, each featuring, along with two binary measurement settings, three preparations; two of them are associated with the certification task, while the third one serves as an auxiliary. The final witness is chosen to self-test some target configuration. We numerically derive a bound $Q_{\text{mirror}}$ for any mirror-symmetric configuration, thereby certifying asymmetry if this bound is exceeded (e.g. experimentally) for the unknown qubit configuration. We also consider the gap $(Q_{\text{max}}-Q_{\text{mirror}})$ between the analytically derived overall quantum maximum $Q_{\text{max}}$ and the mirror-symmetric bound, and use it as a quantifier of asymmetry in the target configuration. Numerical optimization shows that the most asymmetric configuration then forms a right scalene triangle on the unit Bloch sphere. Finally, we implement our protocol on a public quantum processor, where a clear violation of the mirror-symmetric bound certifies asymmetry in the configuration of our experimental triple of qubit states.
\end{abstract}

\maketitle

\section{Introduction}\label{sec:intro}

A typical quantum experiment begins with the preparation of a quantum state and ends with its measurement. Recent advances in black-box quantum information raise the question~\cite{Brunner2014Bell,Pironio2016}: what can be inferred from the measurement statistics when the state preparation device (assigned to one experimenter) and the measurement device at a distance (assigned to another experimenter) are uncharacterized, assuming only minimal assumptions about the systems? In general, the purpose of certification is to verify that a quantum device is functioning correctly, i.e., works according to its specification~\cite{Eisert2020}.
Remarkably, one can certify properties in a black-box manner, such as the system's minimal dimension~\cite{Brunner2008Testing}, entanglement~\cite{Bowles2018}, the existence of a complex Hilbert space~\cite{Renou2021}, the amount of randomness~\cite{Acin2016Certified}, or even the type of measurements performed~\cite{Rabelo2011}. These tasks are collectively known as quantum certification~\cite{Supic2020,Eisert2020}.  

In the dimension-bounded prepare-and-measure (PM) scenario, one experimenter prepares quantum states of a given dimension $D$ and sends the quantum message to another who performs measurements (see e.g. Refs.~\cite{Gallego2010,Pawlowski2011, Bowles2015,Tavakoli2018}). 
Such semi-device independent tasks find application in quantum information processing protocols like quantum key distribution (QKD)~\cite{Pawlowski2011,Chaturvedi2018}, randomness certification~\cite{Li2012,Passaro2015}, quantum random access codes~\cite{Ambainis1999,Ambainis2009,Alves2023}, and self-testing~\cite{Tavakoli2020,Tavakoli2020b,Navascues2023,Drotos2024}.

In this work we develop a method to certify the absence of mirror symmetry in a ``set'' (configuration) of different prepared states within the PM paradigm and introduce an alternative quantifier of asymmetry in the configuration of three Bloch vectors (called target in what follows). Our method is based on constructing a witness that self-tests the (generally asymmetric) target. In particular, any experimentally obtained value of this witness that exceeds the maximum one attainable by a symmetric configuration will certify asymmetry. This kind of approach to certify some qualitative property even in the presence of experimental noise has been in use since~\cite{Tavakoli2018} at least and has explicitly been described in~\cite{Drotos2024}.

In physics, symmetry means the property of a physical system that it is preserved or remains unchanged under a (continuous or discrete) transformation~\cite{lederman2011}. It is often formalized in terms of group representations, and also plays a crucial role in quantum information. Here, we focus on mirror symmetry (or reflection invariance) for qubit states.

An arbitrary qubit state labeled by $x$ can be specified by a density matrix
\begin{equation}
\label{rhox}
\rho_x=\frac{\id_2+\vec n_x\cdot\vec\sigma}{2},    
\end{equation}
where the Bloch vector $\vec n_x$ with $|\vec n_x|\le 1$ completely determines the state, and $\vec{\sigma}$ is the Pauli vector $(\sigma_{x}, \sigma_{y}, \sigma_{z})$. Considering a configuration of three such states ($x=1,2,3$), their Bloch vectors $\vec n_x$ form the vertices of a triangle on the Bloch sphere. A triangle with a reflection axis (i.e., an isosceles triangle) has mirror symmetry. The certification procedure in this study excludes such configurations using a method based on linear witnesses.

Specifically, we modify the standard $I_3$ witness in Ref.~\cite{Gallego2010}, able to self-test three preparations and two binary measurements in a qubit PM setting, by introducing a bias parameter $\omega$ to form a one-parameter family $I_3(\omega)$. By combining three biased witnesses $I_3(\omega_{ij})$ for pairs $(ij)=(12),(13),(23)$, we construct a new witness $I_6$. Here, $(ij)$ refers to a specific pair of preparations, which in the noiseless case, are self-tested using $I_3(\omega_{ij})$. The set of pure state preparations with triple unit Bloch vectors corresponds to the vertices of a triangle in the Bloch representation. Simultaneously self-testing all three vertex pairs of the triangle with the $I_6$ witness, in turn, self-tests the entire triangle.

In the experimentally relevant noisy case, a perfect self-test is not possible. However, the key idea is that if an actual value of $I_6$ (e.g., one obtained in an experiment) exceeds the bound attainable by mirror-symmetric configurations, then the configuration of the Bloch vectors must be asymmetric. Note that this is only possible if the target configuration, which would be self-tested by the ideal value of $I_6$, is asymmetric. The difference between the ideal value of $I_6$ (denoted by $Q_{\text{max}}$ and corresponding to the optimum among all possible configurations, without any constraint, as permitted by quantum mechanics; also known as the ``quantum bound'') and the mirror-symmetric bound (which is obtained by optimizing for $I_6$ over all mirror-symmetric configurations and which we denote by $Q_{\text{mirror}}$) will serve as our quantifier of asymmetry; we will call it the witness gap and denote it by $\Delta$, $\Delta = Q_{\text{max}}-Q_{\text{mirror}}$. Unlike traditional geometric quantifiers of asymmetry (see e.g. Refs.~\cite{Buda1991,Buda1992,Weinberg1993,Petitjean1997}), our approach relies on laws of physics.

The paper is organized as follows. In Sec.~\ref{sec:pre}, we review distance measures and define (lack of) mirror symmetry. Sec.~\ref{sec:pmwit} introduces the qubit PM scenario. Sec.~\ref{sec:I6} details our modification of the standard $I_3$ witness, which leads to the biased $I_3(\omega)$ witness, with the corresponding self-testing properties, then it presents the construction of our actual $I_6$ witness. Sec.~\ref{sec:asymmetry} presents our principles of asymmetry certification, as well as an efficient computation and a quantitative analysis of the mirror-symmetric bound $Q_{\text{mirror}}$ along with the corresponding witness gap $\Delta$. The same section also reports on the experimental certification of asymmetric configurations on a publicly accessible quantum processor. Sec.~\ref{sec:conc} concludes the paper with a summary of the results and a brief outlook on the certification of further asymmetries in quantum systems.

\section{Preliminaries}\label{sec:pre}

A qubit state $\rho_x$ is defined by its density matrix, as given in Eq.~(\ref{rhox}). Here, $\vec n_x$ denotes the Bloch vector corresponding to $\rho_x$, which satisfies $|\vec n_x|\le 1$. For a pure state, $|\vec n_x|= 1$. In our study, we consider geometrical relations between the Bloch vectors of three arbitrary qubit states $\rho_x$ labeled by $x=1,2,3$. 

\subsection{Distance measure}

The trace distance between two quantum states \(\rho_1\) and \(\rho_2\) is defined as
\begin{equation}
D(\rho_1,\rho_2)=\frac{1}{2}\|\rho_1-\rho_2\|_1=\frac{1}{2}\Tr\sqrt{(\rho_1-\rho_2)^\dagger(\rho_1-\rho_2)}.\nonumber    
\end{equation}
For qubit states expressed in the Bloch sphere representation (see Eq.~(\ref{rhox})), this distance simplifies to
\begin{equation}
D(\rho_1,\rho_2)=\frac{1}{2}|\vec n_1-\vec n_2|.
\end{equation}
This measure is invariant under any (anti-)unitary transformation $U$, meaning
\begin{equation}
D(U\rho_1U^\dagger,U\rho_2U^\dagger) = D(\rho_1,\rho_2).    
\end{equation}
In the Bloch picture, any such transformation corresponds to an orthogonal transformation $R$ (i.e., a rotation, reflection, or their combination), so that
\begin{equation}
D(\rho_1,\rho_2)=\frac{1}{2}|\vec n_1-\vec n_2|=\frac{1}{2}|R\vec n_1-R\vec n_2|.
\end{equation}
Moreover, the trace distance is symmetric: $D(\rho_1,\rho_2)=D(\rho_2,\rho_1)$.

\subsection{Mirror symmetry of three qubits}

We say that the three qubit states \(\rho_x\) (with \(x=1,2,3\)) possess mirror symmetry if, for some permutation \((i,j,k)\) of \(\{1,2,3\}\), the following holds:
\begin{equation}
\label{defmirrorsym}
D(\rho_i,\rho_j)=D(\rho_i,\rho_k)
\end{equation}
or equivalently in terms of Bloch vectors,
\begin{equation}
|\vec{n}_i-\vec{n}_j| = |\vec{n}_i-\vec{n}_k|.    
\end{equation}
In other words, mirror symmetry in the configuration of the Bloch vectors $\vec n_x$ exists if at least one of the relations is satisfied:
\begin{align}
\label{mirrorsym}
|\vec n_1-\vec n_2| &= |\vec n_1-\vec n_3|\nonumber\\
|\vec n_1-\vec n_2| &= |\vec n_2-\vec n_3|\nonumber\\
|\vec n_1-\vec n_3| &= |\vec n_2-\vec n_3|.
\end{align}
If none of these equations hold, the configuration is said to be asymmetric (i.e., lacking mirror symmetry). Geometrically, the three Bloch vectors form the vertices of a triangle on the Bloch sphere. Mirror symmetry corresponds to the triangle being isosceles (or equilateral), whereas a scalene triangle (with all sides unequal) indicates asymmetry. See, however, the corresponding discussion in the concluding section.

\section{Witnesses in the prepare-and-measure scenario}\label{sec:pmwit}

In this section, we discuss linear witnesses in the qubit PM scenario. We start by defining the most general two-outcome qubit measurements and the corresponding probability distributions when measuring a set of quantum states.

\subsection{Qubit POVM measurements}

Any two-outcome qubit POVM labeled by $y$ can be represented by two positive operators $M_{b|y}\succeq 0$ for $b=0,1$ which sum to the identity, $\id_2$. For a rank-1 projective measurement with binary outcomes $b\in\{0,1\}$, the measurement operators are given by rank-1 projectors  
\begin{equation}\label{projrank1}
    M_{b|y} = \frac{\id_2+(-1)^b\vec m_y\cdot\vec\sigma}{2},
\end{equation}
where $\vec{m}_{y}$ is the unit Bloch vector corresponding to the two-outcome projective measurement labeled by $y$. A projective measurement can also be degenerate (i.e., consisting only of rank-0 and rank-2 POVM elements), which can be written as the pairs $M_{0|y}=0$, $M_{1|y}=\id_2$ or $M_{0|y}=\id_2$, $M_{1|y}=0$. These are the only extremal two-outcome POVMs for qubits~\cite{DAriano2005}, and a generic two-outcome qubit POVM can be expressed as a convex combination of these extremal POVMs. 

A two-outcome observable can be written in terms of the above POVM elements as 
\begin{equation}
B_y=M_{0|y}-M_{1|y}.    
\end{equation}
For rank-1 projectors, considering Eq.~(\ref{projrank1}), this gives us $B_y=\vec m_y\cdot\vec\sigma$, whereas for degenerate projectors it yields $B_y=\pm \id_2$. More generally, any two-outcome qubit observable can be expressed as 
\begin{equation}
\label{OBSgeneric}
B_y=c_y\id_2+(1-|c_y|)\vec m_y\cdot\vec \sigma,
\end{equation}
where $|c_y|\le 1$. Note that $c_y=0$ gives a traceless observable $B_y=\vec m_y\cdot\vec\sigma$ corresponding to rank-1 projectors, whereas $B_y=c_y\id_2$ with $c_y=\pm 1$ corresponds to degenerate projectors.

\subsection{Probability distributions}

The probability of observing outcome $b$ when a quantum state $\rho_x$ is measured with a rank-1 projective measurement in setting $y$ is given by the Born rule:
\begin{equation}\label{ProbBorn}
    P(b|x,y) = \Tr{(\rho_x M_{b|y})}=\frac{1 +(-1)^b\vec{n}_x\cdot\vec{m}_y}{2}.
\end{equation}
Then the expectation value is conveniently expressed as
\begin{equation}
E_{xy}=P(b=0|x,y)-P(b=1|x,y)=\vec n_x\cdot\vec m_y.
\end{equation}
For a generic two-outcome observable given in Eq.~(\ref{OBSgeneric}), the expectation value becomes 
\begin{equation}\label{Expectgeneric}
E_{xy}=c_y+(1-|c_y|)\vec n_x\cdot\vec m_y,
\end{equation}
where $c_y=\pm 1$ corresponds to the degenerate cases and $c_y=0$ corresponds to the rank-1 projective case.

Formally, a linear witness $I$ is defined as a linear combination of the expectation values $E_{xy}$: 
\begin{equation}
\label{genericwitness}
I=\sum_{x,y}W_{xy}E_{xy},
\end{equation}
where $W_{xy}$ are the coefficients of the witness (also known as elements of a witness matrix $W$). Our task is to derive bounds on the witness $I$ under various choices of coefficients $W_{xy}$ and various choices of assumptions on $E_{xy}$ defined by Eq.~(\ref{Expectgeneric}).

\section{Construction of the $I_6$ witness}
\label{sec:I6}
In this section, we introduce our proposed prepare-and-measure witness, denoted by $I_6$, which is designed for self-testing a triple of qubit states.

\subsection{Biased $I_3$ witness}

\begin{figure}[!t]
    \centering
    \includegraphics[width=.9\linewidth]{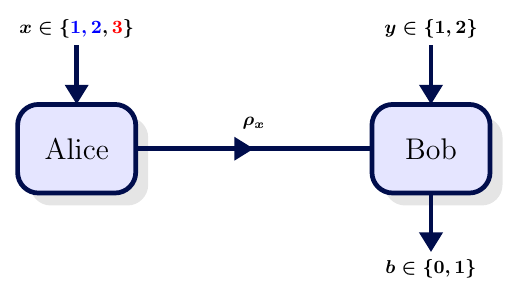}
    \caption{\justifying Prepare-and-measure scenario used in self-testing a pair of qubit states $\rho_1$ and $\rho_2$ along with an auxiliary state $\rho_3$ using two dichotomic measurements. Alice's preparation device randomly generates a state $\rho_x$, $x=1,2,3$, which is sent to Bob. Bob performs a measurement labeled by $y=1,2$ on the received state.}
    \label{fig:1}
\end{figure}

We now specify the form of the $W$ matrix in (\ref{genericwitness}) by defining the biased $I_3(\omega)$ witness, a generalization of the original (unbiased) $I_3$ witness of Gallego et al.~\cite{Gallego2010}, as 
\begin{equation}
\label{I3witness}
I_3(\omega)=\omega(E_{11}+E_{21}-E_{31})+(1-\omega)(E_{12}-E_{22}).   
\end{equation}
This PM witness involves three different state preparations and two dichotomic measurements. See Fig.~\ref{fig:1} for a schematic view of the scenario.

By setting $\omega=1/2$, we recover the original (unbiased) $I_3$ witness. The parameter $\omega\in[0,1]$ in the modified witness gives a bias weight to the measurement settings. For $\omega=0$ and $\omega=1$, only a single measurement setting is used, and in these cases there is known to be no quantum advantage~\cite{Frenkel2015}; i.e., the optimal witness value can be attained by preparations and measurements that can be described classically. We will return to this remark in the concluding section with regard to triangles defined such that two perfectly anti-aligned Bloch vectors are involved.

\begin{figure}[!t]
    \centering
    \includegraphics[width=1\linewidth]{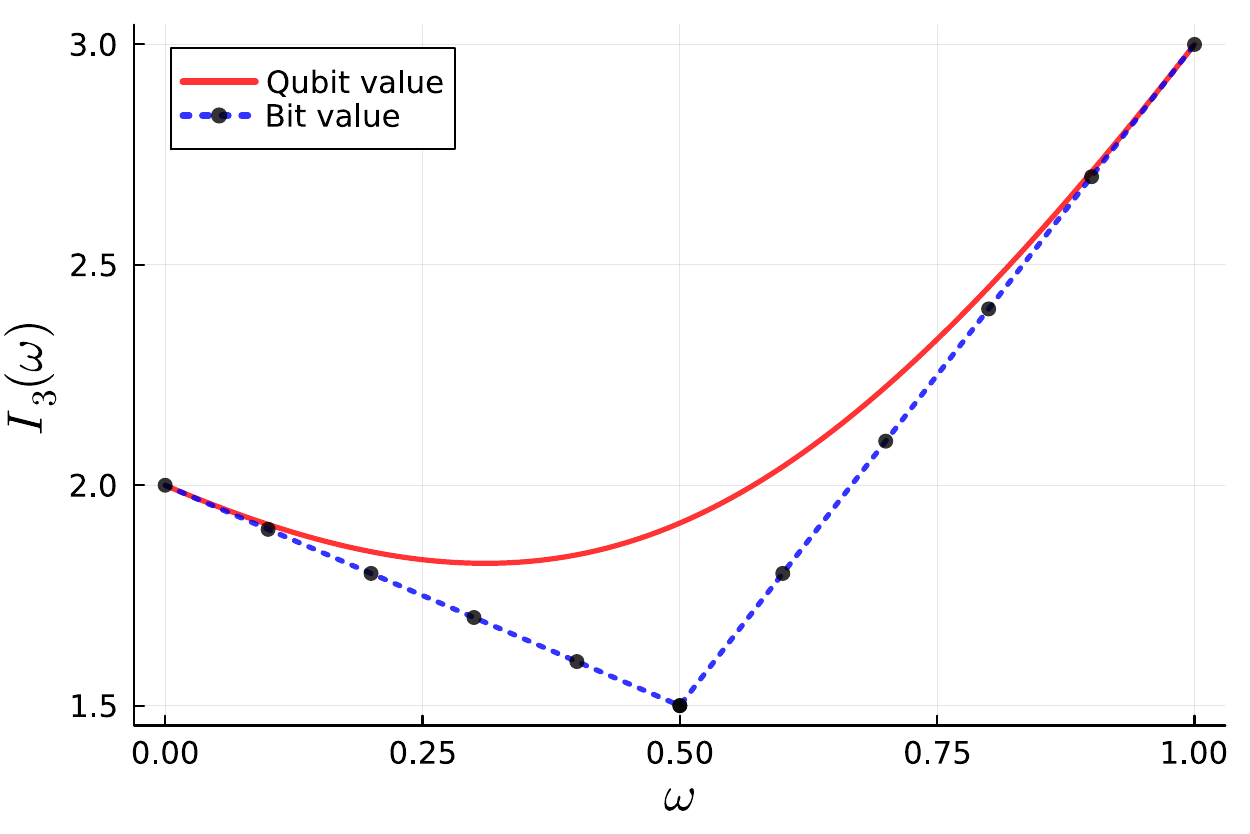}
    \caption{\justifying Maximum quantum value $I_3^Q(\omega)$ and classical bit value $I_3^C(\omega)$ of the witness as a function of the bias parameter $\omega$. The two curves coincide at $\omega=0$ and $\omega=1$, while for intermediate values the quantum maximum exceeds the classical bit value.} 
\label{fig:01}
\end{figure} 

The bound on the classical one-bit value of the witness is a linear function of the bias $\omega$:
\begin{equation}
    I_{3}^{\text{C}}(\omega)=\begin{cases}
         2-\omega,    & \text{if } 0\le\omega\le \frac{1}{2},  \\[1mm]
         3\omega, & \text{if } \frac{1}{2}\le\omega\le 1.
    \end{cases}
\end{equation}
This bound can be achieved with specific deterministic one-bit strategies. While this bound is analytic, efficient numerical methods also exist to compute the one-bit bound of such linear witnesses, even for many settings~\cite{Marton2025}.

In contrast, the maximum quantum value for a given $\omega$ is 
\begin{equation}
\label{I3Qmax}
I_3^{\text{Q}}(\omega)=2\sqrt{\omega^2+(1-\omega)^2}+\omega.
\end{equation}
Fig.~\ref{fig:01} shows both the classical (bit) and quantum (qubit) bounds.

To prove that Eq.~(\ref{I3Qmax}) is the qubit maximum, note that due to the linearity of the witness~(\ref{I3witness}) it suffices to consider pure qubit states (with unit Bloch vectors $\vec n_x$) and projective measurements. Moreover, the $I_3(\omega)$ witness is such that the quantum maximum can be achieved with traceless observables $B_y$ (i.e., with unit Bloch vectors $\vec m_y$ and $c_y=0$ in (\ref{OBSgeneric}) unless $\omega=0$ or $\omega=1$ is chosen; however, the measurement setting that is actually kept must still correspond to a traceless observable in these cases). See Appendix~\ref{app:lemma1} for a proof. Then we have
\begin{equation}
I_3^Q(\omega) = \max_{\vec m_1,\vec m_2}{(|\vec u_+|+|\vec u_-|)+\omega}, 
\end{equation}
where
\begin{equation}
\vec u_{\pm}=\omega\vec m_1 \pm (1-\omega)\vec m_2.
\end{equation}
Applying the Cauchy-Schwarz inequality, we obtain the (tight) bound
\begin{equation}
I_3^Q \le \sqrt{2}\sqrt{|\vec u_+|^2+|\vec u_-|^2}+\omega= 2\sqrt{\omega^2+(1-\omega)^2}+\omega, 
\label{I3Qbound}
\end{equation}
with equality when $\vec m_1\cdot \vec m_2=0$. Besides this, for any $0\leq\omega\leq1$ the value $I_3^Q(\omega)$ self-tests the relation $\vec n_1\cdot\vec n_2$ between the pure state preparations as well, as expressed later in Eq.~(\ref{anglen1n2}). This can be shown with the methodology of Refs.~\cite{Alves2023,Drotos2024Self}, using similar analytic techniques as therein.

The bound in~(\ref{I3Qbound}) is tight, as it can be realized with the following pure qubit preparations: 
\begin{equation}\label{optprepI3}
    \begin{aligned}
    & \vec n_1 =  \frac{\omega \vec m_{1} + (1-\omega) \vec m_{2}}{\sqrt{(\omega^{2}+(1-\omega)^2}},\\
    & \vec n_2 = \frac{\omega \vec m_{1} - (1-\omega) \vec m_{2}}{\sqrt{(\omega^{2}+(1-\omega)^2}},\\
    & \vec n_3 =-\vec m_1,  
    \end{aligned}
\end{equation}
where $\vec m_1$ and $\vec m_2$ are the Bloch vectors corresponding to two mutually orthogonal rank-1 measurements, satisfying $|\vec m_1|=|\vec m_2|=1$ and $\vec m_1\cdot\vec m_2=0$. Note that the cosine of the angle between the unit Bloch vectors $\vec n_1$ and $\vec n_2$ for the optimal preparations is given by
\begin{equation}
\label{anglen1n2}
\cos\alpha=\vec n_1\cdot\vec n_2=\frac{2\omega-1}{\omega^2+(1-\omega)^2}.
\end{equation}
This can be solved for $\omega$: 
\begin{equation}
\omega = \frac{1+\cos\alpha \pm \sqrt{1-\cos^2\alpha}}{2\cos\alpha}    
\label{eq:11}
\end{equation}
for $\cos\alpha\neq 0$. In the special case when $\cos\alpha = 0$ we obtain $\omega=1/2$.

While we have shown here what the quantum maximum is, self-testing considerations are more delicate and are discussed in Appendix~\ref{app:lemma1}. The dot product $\vec n_1\cdot\vec n_2$ will always be self-tested along with one of the measurement settings (characterized by $\vec m_1$ or $\vec m_2$). At the same time, the other measurement setting and $\vec n_3$ may remain unconstrained, although only for $\omega = 0$ or $1$. For $0<\omega<1$, all preparations and measurement settings take part of the self-test.

Let us observe that the function $\vec n_1\cdot\vec n_2$ increases monotonically from $-1$ to $+1$ as $\omega$ varies from 0 to 1. For instance, when $\omega=1/2$, which corresponds to the standard unbiased witness $I_3$, the value of $\vec n_1\cdot\vec n_2$ is $0$, indicating orthogonal preparation vectors (i.e., $\cos\alpha=0$). This dependence gives rise to a remarkable opportunity: should one wish to self-test \emph{any} pair of pure qubit preparations, it becomes viable by an appropriate choice of $\omega$ (according to (\ref{eq:11})) and using a third, auxiliary state in the protocol (which, however, cannot be chosen at will but must satisfy (\ref{optprepI3}) except for the special case when it is not actually required).

\subsection{Self-testing three pure qubits}

\begin{figure}[!t]
    \centering
\includegraphics[width=.9\linewidth]{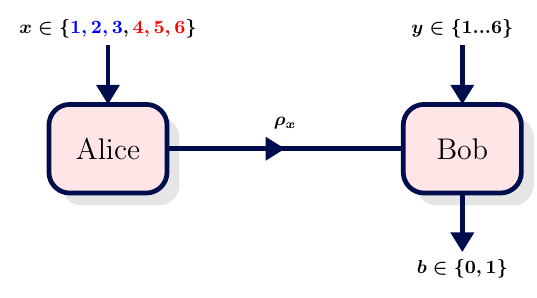}
    \caption{\justifying Schematic of the prepare-and-measure scenario illustrating the self-test and certification of the asymmetry of a triple of qubit states. Alice's preparation device randomly generates a qubit state $\rho_x$, chosen from six states labeled by $x=1,\ldots,6$. She then sends the state to Bob, who performs a measurement labeled by $y$ on the received qubit. The inputs $x = 4,\ldots,6$ correspond to auxiliary qubits required for our scenario to self-test the three qubit states labeled by $x = 1,\dots,3$.} 
    \label{fig:PM_scenario}
\end{figure}

To self-test a triple of qubit states in our scenario (see Fig.~\ref{fig:PM_scenario}), we simultaneously self-test all three pairs of qubits $(12)$, $(13)$ and $(23)$ using the witness $I_3(\omega)$ as a building block. This requires two measurement settings per pair, resulting in a total of six settings. Additionally, three auxiliary preparations are needed, which gives a total of six state preparations.

For the self-test, we first redefine the biased witness $I_3(\omega)$ from Eq.~(\ref{I3witness}) as 
\begin{align}
I_3(\omega; i_1,i_2,i_3,j_1,j_2)\equiv&\omega(E_{i_1,j_1}+E_{i_2,j_1}-E_{i_3,j_1})\nonumber\\
&+(1-\omega)(E_{i_1,j_2}-E_{i_2,j_2}).
\label{I3witij}
\end{align}
Here indices $i_1$, $i_2$ and $i_3$ refer to three of the six state preparations, while $j_1$ and $j_2$ correspond to a pair of measurement settings out of the six possible settings. With this definition, we obtain the $6\times 6$ witness as
\begin{align}
I_{6}&(\omega_{12},\omega_{13},\omega_{23})\equiv
I_3(\omega_{12};1,2,4,1,2)\nonumber\\&+I_3(\omega_{13};1,3,5,3,4)+I_3(\omega_{23};2,3,6,5,6).   \label{I6witness}
\end{align}
This witness can also be expressed in matrix form: 
\begin{equation}
I_{6}(\omega_{12},\omega_{13},\omega_{23})=\sum_{x=1}^6\sum_{y=1}^6 W_{xy}E_{x,y},
\label{I6witnessmat}
\end{equation}
where the matrix $W$ is
\begin{equation*}
\begin{pmatrix*}[r]
& \omega_{12}  & 1-\omega_{12}   & \omega_{13} & 1-\omega_{13}  & 0 & 0 \\
& \omega_{12}  & -1+\omega_{12}   & 0   & 0 & \omega_{23} & 1-\omega_{23}  \\
&  0  & 0   & \omega_{13}  & -1+\omega_{13}  & \omega_{23}   & -1+\omega_{23}  \\ 
& -\omega_{12}  & 0   & 0 & 0  & 0 & 0 \\
& 0  & 0   & -\omega_{13}   & 0 & 0 & 0  \\
& 0  & 0   & 0  & 0  & -\omega_{23}  & 0  \\ 
\end{pmatrix*}
.
\end{equation*}
Next, we compute the maximum qubit bound $Q_{\text{max}}$ for (\ref{I6witness}) and show its self-testing property. Obviously, $Q_{\text{max}}$ is upper bounded by the sum of the quantum maxima of the $I_3(\omega_{ij})$ witnesses in Eq.~(\ref{I3Qmax}), yielding
\begin{equation}
\label{qmax}
    Q_{\text{max}}\le\sum_{ij\in\{(12),(13),(23)\}}2\sqrt{\omega_{ij}^2+(1-\omega_{ij})^2}+\omega_{ij}.
\end{equation}
We now set the three target pure states with specific unit Bloch vectors $\vec{n}^*_1$, $\vec{n}^*_2$, and $\vec{n}^*_3$. Also, we assume that the Bloch vectors are not all collinear and none of them are perfectly aligned, i.e., $\vec n_i^*\cdot\vec n_j^*=-1$ for at most one pair $(i,j)$ and $|\vec n_i^*\cdot\vec n_j^*|<1$ for the rest. Then we choose the parameters $\omega_{12}$, $\omega_{13}$, and $\omega_{23}$ in the witness $I_6$ according to the relation~(\ref{eq:11}). This constitutes a self-testing statement: with this particular choice, every $I_3$ term in the $I_6$ witness~(\ref{I6witness}) can be maximized, thereby saturating the bound in~(\ref{qmax}). Upon choosing the target vectors $\vec{n}^*_1$, $\vec{n}^*_2$, and $\vec{n}^*_3$ such that they lack mirror symmetry, this result can be utilized for a certification of the absence of mirror symmetry also in three Bloch vectors of some unknown configuration. This is elaborated on in the following section.

\section{Certifying asymmetry}
\label{sec:asymmetry}

\subsection{Theory}

It is shown here that the inequality 
\begin{equation}
I_6(\omega_{12},\omega_{13},\omega_{23}) \le Q_{\text{mirror}}    
\end{equation}
serves as a witness. In particular, any violation of this bound for a fixed set of parameters $(\omega_{12},\omega_{13},\omega_{23})$ indicates that the triangular configuration of the Bloch vectors of the prepared states is asymmetric, i.e., it lacks mirror symmetry. Note that this result holds for arbitrary three qubit states; importantly, it does not assume pure state preparations. On the other hand, a single choice of $\{\omega_{12},\omega_{13},\omega_{23}\}$ is not sufficient for certifying the asymmetric nature of an arbitrary asymmetric configuration. Our claim is about the existence of such configurations.

At the same time, choosing the witness appropriately, through the parameters $\{\omega_{12},\omega_{13},\omega_{23}\}$, enables this certification for an arbitrary asymmetric configuration. As shown in the previous section, any target asymmetric pure triple of qubits can be self-tested by an $I_6$ witness, which implies a violation of the mirror-symmetric bound of the given $I_6$ witness for that triple. More generally, as proved in Ref.~\cite{Navascues2023}, any ensemble of pure quantum states with dimension $D\ge 2$ can be self-tested in a PM scenario, including a triple of pure qubits as a special case, even though the witness construction may be different from our $I_6$ witness.

How do we define the mirror-symmetry bound $Q_{\text{mirror}}$? This is the maximum value of the witness $I_6(\omega_{12},\omega_{13},\omega_{23})$ over all mirror-symmetric configurations (i.e., over the set of isosceles and equilateral triangles inscribed in the Bloch sphere). The actual computation of $Q_{\text{mirror}}$ involves an optimization over three different mirror-symmetric configurations, as defined by the Bloch vectors in Eq.~(\ref{mirrorsym}), each of which may yield a different bound for the witness. 

Let us now give a mathematical formulation of the optimization problem for obtaining $Q_{\text{mirror}}$. First, we fix the target states with unit Bloch vectors $\vec{n}^*_{x}$ for $x=1,2,3$, which in turn fixes the parameters of $\omega_{12}$, $\omega_{13}$, and $\omega_{23}$ according to Eq.~(\ref{eq:11}); assumptions are as in the last paragraph of Section~\ref{sec:I6} so as to define a non-degenerate triangle. Our goal is to maximize the witness $I_6$ given in Eq.~(\ref{I6witness}) over all (possibly mixed) qubit states (see Eq.~(\ref{rhox})) satisfying the symmetry conditions in Eq.~(\ref{mirrorsym}) and over all $\pm 1$-valued observables $B_y$ defined by Eq.~(\ref{OBSgeneric}).

However, some simplifications occur at this point. Due to Lemma~\ref{lemma1} of Appendix~\ref{app:lemma1}, for a given set of Bloch vectors $\vec{n}_x$, with $x=1,2,3$, the maximum value of $I_3$ can be achieved if (although not only if) all measurement operators are chosen to be traceless, corresponding to observables of the form $B_y = \vec{m}_y \cdot \vec{\sigma}$. Furthermore, in the optimal case, $\vec{n}_4$ anti-aligns with $\vec{m}_1$, $\vec{n}_5$ anti-aligns with $\vec{m}_3$, and $\vec{n}_6$ anti-aligns with $\vec{m}_5$, according to (\ref{optprepI3}) after appropriate replacement of indices. Since $\vec{n}_4$, $\vec{n}_5$ and $\vec{n}_6$ appear in separate $I_3(\omega; i_1,i_2,i_3,j_1,j_2)$ terms within $I_6$ and also in a single individual term, independent of the rest of the state Bloch vectors and the other measurement Bloch vector, within each $I_3(\omega; i_1,i_2,i_3,j_1,j_2)$, and since they are not involved in symmetry considerations, these anti-alignments carry over to the symmetry-constrained optimum of $I_6$. Consequently, we arrive at the following (nonconvex) optimization problem:
\begin{align}
\label{Qmirror_sdp}
    &Q^{ijk}_{\text{mirror}} = \max \Bigl( \omega_{12} + \omega_{13} + \omega_{23} \nonumber\\
    &+ \omega_{12}\,(\vec{n}_1+\vec{n}_2)\cdot \vec{m}_1 + (1-\omega_{12})\,(\vec{n}_1-\vec{n}_2)\cdot\vec{m}_2 \nonumber\\ 
    &+ \omega_{13}\,(\vec{n}_1+\vec{n}_3)\cdot \vec{m}_3 + (1-\omega_{13})\,(\vec{n}_1-\vec{n}_3)\cdot\vec{m}_4 \nonumber\\ 
    &+ \omega_{23}\,(\vec{n}_2+\vec{n}_3)\cdot \vec{m}_5 + (1-\omega_{23})\,(\vec{n}_2-\vec{n}_3)\cdot\vec{m}_6 \Bigr),
\end{align}
where the maximization is taken over all measurement vectors $\vec{m}_y \in S^2$ for $y=1,\ldots,6$ and preparation Bloch vectors $\vec{n}_x \in \mathbb{R}^3$ for $x=1,2,3$ satisfying $|\vec{n}_x|\leq 1$, and, subject to the additional constraint
\begin{equation}
|\vec{n}_i-\vec{n}_j| = |\vec{n}_i-\vec{n}_k|,
\label{nijik}
\end{equation}
for a given permutation $(i,j,k)$ of the set $\{1,2,3\}$, according to the mirror symmetry conditions in Eq.~(\ref{mirrorsym}). Note that in Eq.~(\ref{Qmirror_sdp}) there is no loss of generality in assuming unit Bloch vectors $\vec m_y$, for $y=(1,\ldots,6)$.

There are three separate mirror-symmetry conditions, which yield the bounds $Q_{\text{mirror}}^{123}$, $Q_{\text{mirror}}^{213}$, and $Q_{\text{mirror}}^{312}$. We solve each case independently and then select the maximum value, which we denote by $Q_{\text{mirror}}$:
\begin{equation}
\label{maxQmirror}
    Q_{\text{mirror}} = \max\{Q_{\text{mirror}}^{123},\, Q_{\text{mirror}}^{213},\, Q_{\text{mirror}}^{312}\}.
\end{equation}

The optimization problem of each $Q_{\text{mirror}}^{ijk}$ term belongs to the class of quadratically constrained quadratic programming (often framed as QCQP)~\cite{Boyd2004}. To see this, note that by squaring the norm in Eq.~(\ref{nijik}) we obtain the quadratic constraint $\vec n_i\cdot\vec n_j=\vec n_i\cdot\vec n_k$. Each Bloch vector $\vec{n}_x = (n_{x1}, n_{x2}, n_{x3})$ for $x=1,2,3$ contributes three scalar variables, and each measurement vector $\vec{m}_y = (m_{y1}, m_{y2}, m_{y3})$ for $y=1,\ldots,6$ also contributes three scalar variables. We also have the constraints $\vec{n}_x \cdot \vec{n}_x \le 1$ and $\vec{m}_y \cdot \vec{m}_y = 1$. 

Solving the optimization problem in Eq.~(\ref{Qmirror_sdp}) is generally computationally still demanding because it may have several local maxima. Following the approach of Lasserre~\cite{Lasserre2001}, we derived an upper bound on $Q_{\text{mirror}}$ at the first level of the hierarchy using semidefinite programming techniques~\cite{Vandenberghe1996}. For each case considered, the computed upper bound matched, within numerical precision, with the maximum value obtained using a heuristic lower-bound method. Thus, our upper bound achieved on the first relaxation level of the hierarchy is tight for all the studied cases. 

Furthermore, the non-convex problem~\eqref{Qmirror_sdp} can be reformulated through variable changes to form a conic programming problem, which guarantees the existence of global tight bounds $Q^{ijk}_{\text{mirror}}$ (see Appendix~\ref{app:purityproof}). Moreover, based on this reformulation, we have proved that the upper bounds $Q_{\text{mirror}}$ and $Q_{\text{max}}$ can only be achieved by pure states (see Lemmas~\ref{lemma2} and ~\ref{lemma3}, and their respective proofs in Appendix~\ref{app:purityproof}).

Besides the mirror-symmetry bound $Q_{\text{mirror}}$ itself (in Eq.~(\ref{maxQmirror})), it may also be relevant how much the ideal value, i.e., the overall physical (quantum mechanical) bound $Q_{\text{max}}$ (see Eq.~(\ref{qmax})) is higher. We thus introduce the witness gap $\Delta$ to characterize this difference. It will serve as a quantifier of asymmetry in our certification scenario and is defined by
\begin{equation}\label{eq:delta}
    \Delta = Q_{\text{max}}(\omega_{12},\omega_{13},\omega_{23}) - Q_{\text{mirror}}(\omega_{12},\omega_{13},\omega_{23}).
\end{equation}
Note that a value of $\Delta > 0$ opens the opportunity to certify asymmetry in the triangular configuration of some unknown preparation Bloch vectors.

\subsection{Quantitative findings for the gap}

\begin{table*}[!t]
\centering
\begin{tabular}{|l l l| c c c|ccc|cc|c|}
\hline
$\alpha^*_{12}$&$\alpha^*_{13}$ &$\alpha^*_{23}$ & 
$\omega_{12}$&$\omega_{13}$ &$\omega_{23}$ &
$Q_{\text{mirror}}^{123}$ & $Q_{\text{mirror}}^{213}$ & $Q_{\text{mirror}}^{312}$ & 
$Q_{\text{max}}$& $Q_{\text{mirror}}$& $\Delta$\\ 
\hline
\hline
130$^\circ$ & 130$^\circ$ & $100^\circ$ & $0.318$&$0.318$ & $0.456$ &$5.52185$& $5.49684$& $5.49684$& $5.52185$&$5.52185$&  $0.00000$\\ \hline
58.4$^\circ$ & 121.6$^\circ$ & $180^\circ$ &$0.641$&$ 0.358 $ &$ 0.000 $ & $5.82843$& $5.46650$& $5.82843$& $5.93950$& $5.82843$&  $0.11107$\\ \hline
54$^\circ$ & 112$^\circ$ & $194^\circ$ &$0.662$&$ 0.403$ &$ 0.109 $ & $5.80372$& $5.51644$& $5.80866$& $5.89696$& $5.80866$&  $0.08831$\\ \hline
\end{tabular}
\caption{\justifying Comparison between the overall bound $Q_{\text{max}}$ and the mirror-symmetric bound $Q_{\text{mirror}}$ on the witness $ I_6(\omega_{12},\omega_{13},\omega_{23})$ defined in Eq.~(\ref{I6witness}) for a given set of angles \(\{\alpha^*_{12}, \alpha^*_{13}, \alpha^*_{23}\}\). The value $Q_{\text{mirror}}$, as defined in Eq.~(\ref{maxQmirror}), is the maximum among the three mirror-symmetric cases, denoted by $Q_{\text{mirror}}^{ijk}$ with $ijk=(123)$, $(213)$, and $(312)$. Three exemplary target configurations are presented. The first row corresponds to preparations that form an isosceles triangle, which is a mirror-symmetric configuration with $\Delta=0$. The second row shows the witness values for the most asymmetric configuration as characterized by the largest gap $\Delta_{\text{max}} = 0.11107$. The last row corresponds to a configuration close to optimal, with all entries given in integer degrees.}
\label{table:1}
\end{table*}

In Table~\ref{table:1}, the first row illustrates a symmetric triangle, where the target Bloch vectors $\{\vec{n}^*_x\}$ form an isosceles triangle with equal angles $\alpha^*_{12} = \alpha^*_{13} = 130^\circ$. These angles determine the biases $\{\omega_{12}, \omega_{13}, \omega_{23}\}$ in the witness $I_6$ (see Eq.~(\ref{I6witness})) through Eq.~(\ref{anglen1n2}). In this case, the mirror-symmetric bound reaches the quantum maximum, so that $\Delta = 0$.

\begin{figure}[htbp]
	\centering
	\begin{subfigure}{0.52\linewidth}
		\includegraphics[width=\linewidth]{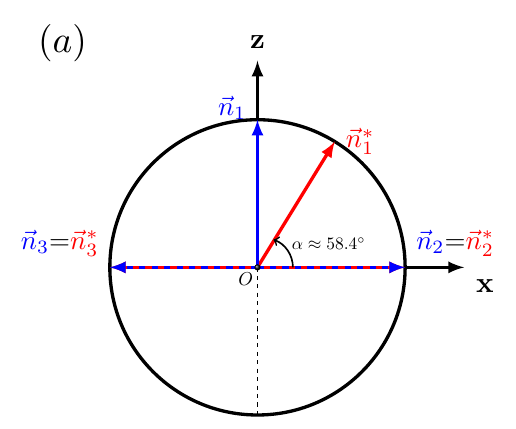}
		\label{fig:expA}
	\end{subfigure}
	\hfill
	\begin{subfigure}{0.45\linewidth}
        \includegraphics[width=\linewidth]{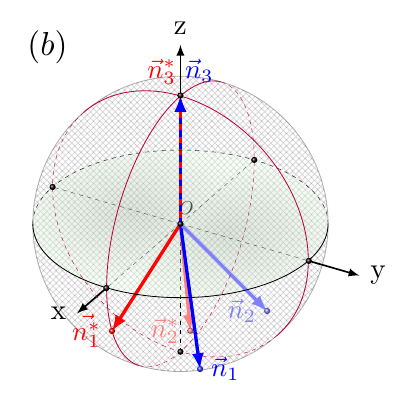}
		\label{fig:expB}
	\end{subfigure}
	\caption{\justifying Illustration of the two asymmetric configurations aimed to be implemented on IBMQ with a success in asymmetry certification. For each case, the corresponding optimal symmetric configuration saturating the bound $Q_{\text{mirror}}$ is also depicted. In the plots, the starred Bloch vectors represent the target asymmetric configuration, while the unstarred ones represent the optimal symmetric configuration. Panel (a) corresponds to the starred angles given in Eq.~(\ref{angles_case1}). Here the optimal symmetric Bloch vectors lie in the same $X$--$Z$ plane. Panel (b) corresponds to the starred angles given in Eq.~(\ref{angles_case2}). Here, the Bloch vectors of the optimal symmetric configuration span the full three-dimensional space, where $\vec{n}_1$ and $\vec{n}_2$ are mirror images of each other with respect to the $Y$--$Z$ plane, while $\vec{n}_3$ is along the $Z$-axis.}
   	\label{fig:exp}
\end{figure}

Let us next discuss the following two rows. We performed a heuristic optimization of $\Delta$ in (\ref{eq:delta}) over all pure target qubit states, with Bloch vectors $|\vec{n}^*_x| = 1$, $x=1,2,3$, and have found that the most asymmetric configuration of the three Bloch vectors is characterized by the angles 
\begin{equation}
\alpha^*_{12} = 58.4\degree,\quad \alpha^*_{13} = 121.6\degree,\quad \text{and} \quad \alpha^*_{23} = 180\degree,
\label{angles_case1}
\end{equation}
in the sense that it yields the largest gap $\Delta_{\text{max}} \simeq 0.11107$ between $Q_{\text{max}}$ and $Q_{\text{mirror}}$ (see the second row of Table~\ref{table:1}). In this case, the vertices of the symmetric triangle that saturates the bound $Q_{\text{mirror}}=3+2\sqrt 2$ lie on a great circle of the Bloch sphere; note also that two Bloch vectors appear to be perfectly anti-aligned. Finally, in the third row, we have the angles in integer degrees
\begin{equation}
\alpha^*_{12} = 54\degree,\quad \alpha^*_{13} = 112\degree,\quad \text{and} \quad \alpha^*_{23} = 194\degree.    
\label{angles_case2}
\end{equation}
Surprisingly (in view of our ad hoc experience), in this configuration the three Bloch vectors of the optimal symmetric arrangement are not coplanar, i.e., they are not all located on a great circle of the Bloch sphere. Fig.~\ref{fig:exp} illustrates these two configurations~(\ref{angles_case1}) and ~(\ref{angles_case2}), which we also aimed to implement on IBMQ with a successful asymmetry certification, as discussed in the next section.

\subsection{Asymmetry certification on IBM's quantum device}

Table~\ref{Tab_exp} summarizes the experimental results obtained from three experiments with different target values of $\alpha^*_{12}$, $\alpha^*_{13}$ and $\alpha^*_{23}$, and the corresponding witness values $I_6(\omega_{12},\omega_{13},\omega_{23})$ according to the formula~(\ref{eq:11}). The last two rows correspond to the asymmetric examples presented in Table~\ref{table:1}. Using the \textit{ibm brisbane} quantum device, we executed quantum circuits designed to implement the optimal configuration, characterized by the gap $\Delta = Q_{\text{max}}-Q_{\text{mirror}}$, but obviously deviating from the optimal configuration in practice. In the first experiment the witness value does not violate the mirror-symmetric bound. However, it does in the last two experiments. Therefore, although we do not know the configurations we actually managed to implement, we can be certain (up to the semi-DI assumptions) that they are asymmetric.

In the second row, $Q_{\text{mirror}}=3+2\sqrt{2}\simeq5.8284$, and all experiments were conducted with 8192 shots. The scripts for computing the $Q_{\text{mirror}}$ bound for a given set of target Bloch vectors, as well as those designed for performing certification tasks on IBM Quantum backends, are made available on GitHub at \cite{Malek2025}. Additionally, we therein demonstrate how to determine the maximal witness gap $\Delta_{\text{max}}$, which we define to be the most asymmetric configuration, using a heuristic optimization technique.

\begin{table*}[t]
\centering
\begin{tabular}{|l l l|cc|ccc|c|}
\hline
$\alpha^*_{12}$ & $\alpha^*_{13}$ & $\alpha^*_{23}$ & $Q_{\text{mirror}}$ & $Q_{\text{max}}$ & $I_3^{\text{exp}}(\omega_{12})$ & $I_3^{\text{exp}}(\omega_{13})$ & $I_3^{\text{exp}}(\omega_{23})$ & $Q_{\text{exp}}$\\ 
\hline
\hline
60$^\circ$    & 200$^\circ$   & 100$^\circ$  & 5.8065  & 5.8503  & 2.0672 & 1.8622 & 1.8582 & $5.7877 \pm 0.0113$\\ \hline
58.4$^\circ$  & 121.6$^\circ$ & 180$^\circ$  & 5.8284  & 5.9395  & 2.0938 & 1.8220 & 1.9736 & $5.8894 \pm 0.0145$\\ \hline
54$^\circ$    & 112$^\circ$   & 194$^\circ$  & 5.8086  & 5.8970  & 2.1195 & 1.8205 & 1.8947 & $5.8347 \pm 0.0106$\\ \hline
\hline
\end{tabular}
\caption{\justifying Results from three experiments with different target values of $\alpha^*_{ij}$ (with $ij=(12)$, $(13)$, and $(23)$) and the corresponding witness values $I_6(\omega_{12},\omega_{13},\omega_{23})$. The last two rows correspond to the asymmetric examples presented in Table~\ref{table:1}. We used the \textit{ibm brisbane} quantum device to certify asymmetry experimentally. Here, $Q_{\text{exp}}$ and $I_3^{\text{exp}}$ denote the experimental values of the witnesses $I_6$ and $I_3$, respectively. In the first experiment, $Q_{\text{exp}}$ does not violate the mirror-symmetric bound, whereas in the last two experiments it does. The standard deviation corresponding to $Q_{\text{exp}}$ has been computed in accordance with the formulas outlined in Appendix~\ref{app1}.}
\label{Tab_exp}
\end{table*}

\section{Conclusion}\label{sec:conc}

We proposed a protocol to certify asymmetry in an unknown configuration of the Bloch vectors associated with a set of three qubit states in a prepare-and-measure scenario, using a biased $I_3$ witness as a building block. In particular, given a target set of three qubits, we constructed a self-test of the triple by testing each pair individually. We have also developed an efficient method to compute the mirror-symmetric bound of the same witness, which allows us to quantify the asymmetry in a given set of three qubit states. It is important to note that our technique differs from conventional definitions of chirality or axiality (e.g., the ones appearing in Refs.~\cite{Buda1991,Buda1992}. 

Furthermore, symmetry as defined here concentrates on the geometry of the triangle formed by the endpoints of the Bloch vectors: in particular, any configuration forming an isosceles triangle is considered symmetric. However, in case the symmetry plane does not include the origin, the symmetry transformations do not leave the geometry of the qubit configuration invariant, so that the qubit configuration as a quantum mechanical construction is not actually symmetric. This issue would become relevant if an optimal symmetric witness value were achieved by some mixed states, as only mixed states may give rise to symmetry planes that do not include the origin according to Eq.~(\ref{mirrorsym}). However, as shown in our analytical proof in Appendix~~\ref{app:purityproof}, the optimal symmetric witness value of $I_6$ is always achieved by pure states. In total, while Eq.~(\ref{mirrorsym}) is more permissive than what would perhaps be quantum mechanically relevant, this does not appear to affect final conclusions.

To test the robustness of our protocol against experimental noise, we implemented it on an IBM quantum device. Selected qubits yielded statistics that allowed us to violate the mirror-symmetric bound. Our theoretical findings, along with experimental demonstrations, thus prove that it is possible to certify the asymmetric nature of a configuration of a set of three unknown qubit Bloch vectors in a semi-device independent PM scenario.

Note that symmetric configurations of three qubit states have already found applications in quantum key distribution protocols (see, e.g., Ref.~\cite{Renes2004}). For instance, the symmetric configuration of trine qubit states along with their anti-trine states has been used in a basis-free quantum key distribution protocol~\cite{Tabia2011}.

We emphasize here that we have allowed for any triangular configuration as a target of our self-test other than those defining degenerate triangles (i.e., with coinciding sides). In particular, triangles defined through two perfectly anti-aligned Bloch vectors and a third one pointing in any different direction are included in our analysis. Even though a self-test of the former two in itself would show no quantum advantage (cf.~\cite{Frenkel2015}), the self-test of the entire triple will still do so according to the remaining two constituents of the whole witness (which self-test the two remaining choices of a qubit pair out of the triple).

We believe that the certification of a set of qubit states (not limited to trine states) may have diverse applications in black-box quantum communication. Furthermore, we are planning to extend the asymmetry certification to larger sets of qubit states, such as tetrahedral configurations inscribed in the Bloch sphere. It would also be interesting to quantify the asymmetry in measurement Bloch vectors and in higher-dimensional quantum systems.

\begin{acknowledgments}
We acknowledge the support of the EU (CHIST-ERA MoDIC), the National Research, Development and Innovation Office NKFIH (No.~2023-1.2.1-ERA\_NET-2023-00009 and No.~K145927). T.V. acknowledges support from the ``Frontline'' Research Excellence Program of the NKFIH (No.~KKP133827). G.D. acknowledges support from the European Union (European Social Fund Plus) and the Government of the Balearic Islands through a Vicen\c{c} Mut postdoctoral fellowship (No. PD-035-2023).
\end{acknowledgments}

\appendix

\section{Lemma~1 and implications}\label{app:lemma1}

To establish the certification property of $I_3(\omega)$ for state preparations $\rho_1$ and $\rho_2$, we first present the lemma below.

\begin{lemma}\label{lemma1}
Consider the witness $I_3(\omega)$ defined in Eq.~(\ref{I3witness}) with $0 \leq \omega \leq 1$. For any set of state preparations (whether mixed or pure) characterized by the Bloch vectors $\vec{n}_1$ and $\vec{n}_2$, the maximum quantum value is given by
\begin{equation}
I_3^Q(\omega,\vec{n}_1,\vec{n}_2)=\omega\,|\vec n_1 + \vec n_2| + (1-\omega)\,|\vec n_1 - \vec n_2| + \omega,    
\label{I3opt}
\end{equation}
where the optimization is taken over the state $\rho_3$ with Bloch vector $\vec n_3$ and the dichotomic observables $B_y$, for $y=1,2$. When the witness is defined with $0 < \omega < 1$, and both $|\vec n_1 +\vec n_2|>0$ and $|\vec n_1 -\vec n_2|>0$, the maximum~(\ref{I3opt}) can be attained only if traceless observables are used (i.e., $B_y=\vec m_y\cdot\vec\sigma_y$, for $y=1,2$). However, if $\vec n_1 + \vec n_2=0$ or $\vec n_1 - \vec n_2=0$ holds, the maximum~(\ref{I3opt}) can also be achieved if $B_1$ and $B_2$, respectively, describes a degenerate measurement. If $\omega = 0$ or $\omega = 1$ is chosen, then $B_1$ and $B_2$, respectively, remain unconstrained.
\end{lemma}

\noindent\textit{Proof.} --- Our goal is to compute
\begin{align}
\label{Qmax_n1n2}
I_3^Q(\omega,\rho_1,\rho_2) = \max_{B_1,B_2,\rho_3} \Biggl[ & \omega\,\Tr\Bigl(B_1\bigl(\rho_1+\rho_2-\rho_3\bigr)\Bigr) \nonumber\\[1mm]
& + (1-\omega)\,\Tr\Bigl(B_2\bigl(\rho_1-\rho_2\bigr)\Bigr) \Biggr],
\end{align}
i.e., the maximization is performed over the observables $B_1$, $B_2$, and the state $\rho_3$.

First, consider the case of traceless observables,
$B_y = \vec{m}_y\cdot\vec{\sigma}$, for $y=1,2$. In this case, the expression becomes
\begin{align}
I_3^Q(\omega,\rho_1,\rho_2) =& \max_{\vec m_1,\vec m_2,\vec n_3} \Biggl[ \omega\,\vec m_1\cdot\bigl(\vec n_1+\vec n_2\bigr) -\omega\vec m_1\cdot\vec n_3 \nonumber\\
&  + (1-\omega)\,\vec m_2\cdot\bigl(\vec n_1-\vec n_2\bigr) \Biggr];
\end{align}
note that the maximization is now taken over the vectors $\vec m_1$, $\vec m_2$, and $\vec n_3$. By aligning $\vec m_1$ with $\vec n_1+\vec n_2$ and $\vec{m}_2$ with $\vec n_1-\vec n_2$ along with choosing $\vec{n}_3 = -\vec{m}_1$, we obtain the bound~(\ref{I3opt}) stated in the lemma (even if $\omega = 0$ or $1$).

Next, consider degenerate measurements. Suppose we set $B_2 = \pm \mathbb{I}_2$ in (\ref{Qmax_n1n2}). In this case,
\begin{equation*}
\Tr\Bigl(B_2\bigl(\rho_1-\rho_2\bigr)\Bigr) = 0,    
\end{equation*}
which is the same result as that obtained using a traceless observable $B_2$ only when $\vec n_1 = \vec n_2$; it is strictly inferior otherwise. Similarly, if we set $B_1 = \pm\mathbb{I}_2$ in (\ref{Qmax_n1n2}), then
\begin{equation*}
\Tr\Bigl(B_1\bigl(\rho_1+\rho_2-\rho_3\bigr)\Bigr) = \pm 1.    
\end{equation*}
This value attains the bound $|\vec{n}_1+\vec{n}_2| + 1$, which is achieved using a traceless observable $B_1$, only if $\vec{n}_1 = -\vec{n}_2$.

Any of the above considerations about $B_1$ and $B_2$, respectively, remains irrelevant if $\omega = 0$ or $\omega = 1$ is chosen. In these cases, outcome statistics from the corresponding measurement do not enter the witness, so that the corresponding measurement remains unconstrained. As traceless observables always let attaining the maximum, the bound remains as stated in~(\ref{I3opt}).
\qed

For the assessment of self-testing, we need to distinguish between two principally different cases according to Lemma~\ref{lemma1}.

\begin{itemize}
	\item If we choose $\omega \notin \{0,1\}$ according to formula~(\ref{anglen1n2}) after specifying appropriate target vectors $\vec n_1$ and $\vec n_2$, this allows for self-testing all of the involved states and measurement settings as per the main text. Notably, Lemma~\ref{lemma1} does not claim that reaching the maximum value of the witness always requires the use of traceless observables when making such a choice for $\omega$. However, it does claim that traceless observables are required unless the witness (with an already specified $\omega$) is \emph{evaluated} for some $\vec{n}_1 = -\vec{n}_2$ or $\vec{n}_1 = \vec{n}_2$; in the latter cases, the witness value is the same with traceless observables $B_1$ and $B_2$ and with using instead one degenerate measurement appropriately. Now, we know from the main text that such configurations are not optimal for traceless observables under $0 < \omega < 1$; instead, the witness value can be made higher and will reach its maximum for the target configuration, ensuring thus the self-testing property.
	\item The relationships $\vec{n}_1 = -\vec{n}_2$ and $\vec{n}_1 = \vec{n}_2$ can be self-tested by choosing $\omega = 0$ and $1$, respectively. In these cases, the witness simplifies to a single term with only one measurement setting entering ($B_2$ for $\omega = 0$ and $B_1$ for $\omega = 1$). According to Lemma~\ref{lemma1}, this remaining measurement setting still cannot be degenerate at the self-testing witness value so that it will be self-tested along with the two state preparations $\rho_1$ and $\rho_2$. At the same time, the other measurement setting will (obviously) remain fully unconstrained by the witness, as well as $\vec n_3$ if $\omega = 0$. For $\omega = 1$, $\vec n_3 = - \vec n_1 = - \vec n_2$ will take part of the self-test. However, note that the algebraic maximum of the witness can already be attained using any classical bit in this case, provided that the measurement depends on the classical bit value. Hence, in the particular case of $\omega\in\{0,1\}$, the witness cannot distinguish between classical and quantum bits.  
\end{itemize}

\section{Proof of the purity of qubit states achieving the bound $Q_{\rm{mirror}}$}\label{app:purityproof}

\subsection{Convex program associated with $Q$}

Let us consider the witness that serves as the argument of the maximization function on the right-hand side of~\eqref{Qmirror_sdp} in the main text (which is equivalent to the witness~\eqref{I6witnessmat} when taking into account Lemma~\ref{lemma1} of Appendix~\ref{app:lemma1}). We can eliminate the measurement Bloch vectors by optimizing their choices. The optimal  measurement Bloch vectors $\vec{m}^{*}_j$ for $j=1,...,6$ are given by
\begin{equation*}
\begin{aligned}
    &\vec{m}^*_1 = \vec{m}^{12,+}, \quad \vec{m}^*_2 = \vec{m}^{12,-},\\
    &\vec{m}^*_3 = \vec{m}^{13,+}, \quad \vec{m}^*_4 = \vec{m}^{13,-},\\
    &\vec{m}^*_5 = \vec{m}^{23,+}, \quad \vec{m}^*_6 = \vec{m}^{23,-},
\end{aligned}
\end{equation*}
where 
\begin{equation*}
     \vec{m}^{ij,\pm} = (\vec{n}_i \pm \vec{n}_j)/|\vec{n}_i \pm \vec{n}_j|.
\end{equation*}
Consequently, we establish the following inequality:
\begin{equation*}
\begin{aligned}
Q \leq & \omega_{12} + \omega_{13} + \omega_{23} \\ 
& + \omega_{12}|\vec{n}_1 + \vec{n}_2| + (1 - \omega_{12})|\vec{n}_1 - \vec{n}_2|\\
& + \omega_{13}|\vec{n}_1 + \vec{n}_3| + (1 - \omega_{13})|\vec{n}_1 - \vec{n}_3| \\
& + \omega_{23}|\vec{n}_2 + \vec{n}_3| + (1 - \omega_{23})|\vec{n}_2 - \vec{n}_3|, 
\end{aligned}
\end{equation*}

\noindent which can be expressed as:
\begin{equation*}
Q \leq \omega_{12} + \omega_{13} + \omega_{23} + F(\vec{n}_1, \vec{n}_2, \vec{n}_3),
\end{equation*}

\noindent where
\begin{equation}\label{Fnonlinear}
F = \sum_{ij \in \{12, 13, 23\}} \omega_{ij}|\vec{n}_i + \vec{n}_j| + (1 - \omega_{ij})|\vec{n}_i - \vec{n}_j|.   
\end{equation}

In general, the functional $F$ in Eq.~\eqref{Fnonlinear} is non-convex in this form. To facilitate analysis, we will apply the following change of variables:
\begin{equation}\label{udef}
\begin{aligned}
    u^{ij,+} &:= |\vec{n}_i + \vec{n}_j|,\\
    u^{ij,-} &:= |\vec{n}_i - \vec{n}_j|. 
\end{aligned} 
\end{equation}

\noindent Here, $ u^{ij,\pm} \geq 0 $ for all $ ij \in \{12, 13, 23\}$. Therefore, the functional (\ref{Fnonlinear}) can be rewritten in a linear form as:

\begin{equation}\label{Flinear}
F = \sum_{\substack{ij \in \{12, 13, 23\} \\ p \in \{+, -\}}} A_{ij}^{p} u^{ij, p}, 
\end{equation}

\noindent where $A_{ij}^{+} = \omega_{ij}$, $A_{ij}^{-} = (1-\omega_{ij})$ and $A_{ij}^{\pm}\in[0,1]$. Note that $F$ is subject to the constraints  $u^{ij,\pm} \geq 0$ along with 
\begin{equation}\label{quadConst}
(u^{ij,\pm})^2 \leq c_{ii} + c_{jj} \pm 2c_{ij},\\
\end{equation}

\noindent where  $c_{ij} = \vec{n}_i \cdot \vec{n}_j $ represent the elements of the Gram matrix  $G$  corresponding to the three vectors  $\vec{n}_1, \vec{n}_2,$  and  $\vec{n}_3$:
\begin{equation*}
G = \begin{pmatrix}
c_{11} & c_{12} & c_{13} \\
c_{12} & c_{22} & c_{23} \\
c_{13} & c_{23} & c_{33} 
\end{pmatrix}.
\end{equation*}

\noindent The diagonal elements of the matrix $G$ above satisfy $0 \leq c_{ii} \leq 1 $, while the off-diagonal elements satisfy $ -1 \leq c_{ij} \leq 1 $. Additionally, the Gram matrix is positive semidefinite (PSD). 
Equivalence with the original constraints~\eqref{udef} is ensured by the monotonically increasing dependence of $F$ in~\eqref{Flinear}  on $u^{ij,\pm}$. These quadratic constraints are equivalent to second-order cone constraints. 
Our conic optimization problem takes the following form:
\begin{equation}\label{cvxprogram}
\begin{aligned}
    F^{*} = \max_{G,\vec{u}}   \quad & \vec{a}^\top \vec{u} \\
\text{s.t.} \quad 
    & G\succeq 0, \\
    & G_{ii}\geq 0, \\
    & G_{ii}\leq 1, \quad i=1,2,3 \\
    & (u^{ij,\pm})^2 \leq G_{ii}+G_{jj}\pm 2G_{ij}, ij\in\{12,13,23\}\\
    & \vec{u} \succeq 0,
\end{aligned}
\end{equation}
where $\vec{u}\in \mathbb{R}^6$ is the vector grouping the $u^{ij,\pm}$ variables and $\vec{a}^\top = (A_{12}^{+},A_{12}^{-},A_{13}^{+},A_{13}^{-},A_{23}^{+},A_{23}^{-})$. 

\medskip

\noindent\textbf{Mirror-symmetric case.}---Let $F_{ijk}$ be the witness $F$ in Eq.~\eqref{Flinear} under the symmetry constraint $|\vec{n}_i - \vec{n}_j| = |\vec{n}_i - \vec{n}_k|$. Thus, it follows that $F_{ijk}$ can be written as
\begin{align*}
    F_{ijk} =& (\omega_{ij} + \omega_{ik})u^{ij,+} + (2-(\omega_{ij} + \omega_{ik}))u^{ij,-} \nonumber\\ &+ \omega_{jk}u^{jk,+} + (1-\omega_{jk})u^{jk,-} = \vec{a}^{\top}\vec{u},
\end{align*}
where 
\begin{equation}
\vec{a}^{\top} = (\omega_{ij} + \omega_{ik}, 2 - \omega_{ij} - \omega_{ik}, \omega_{jk}, 1-\omega_{jk})    
\end{equation}
and 
\begin{equation}
\vec{u}^\top = (u^{ij,+},u^{ij,-},u^{jk,+},u^{jk,-}).    
\end{equation}
The upper bound of $F_{ijk}$ can be found by solving the following conic program
\begin{equation}
\begin{aligned}
    F^*_{ijk} = \max_{G,\vec{u}}   \quad & \vec{a}^\top \vec{u} \\
\text{s.t.} \quad 
    & G\succeq 0, \\
    & G_{ll}\geq 0, \\
    & G_{ll}\leq 1, \quad l=1,2,3 \\
    & (u^{ij,\pm})^2 \leq G_{ii}+G_{jj}\pm 2G_{ij},  \quad\quad\, \text{[C1]}\\
    & (u^{ij,\pm})^2 \leq G_{ii}+G_{kk}\pm 2G_{ik},  \quad\quad \text{[C2]}\\
    & (u^{jk,\pm})^2 \leq G_{jj}+G_{kk}\pm 2G_{jk},\\
    & \vec{u} \succeq 0.
\end{aligned}
\label{cvxsymm}
\end{equation}
However, arguing for equivalence with the original problem is not as simple in this case as it was for (\ref{cvxprogram}), since [C1] and [C2] cannot be simultaneously saturated unless $G_{ij}=G_{ik}$. This issue will be taken care of in the proof of purity in the next section.

\subsection{Proof of purity of the optimal qubit states reaching the upper bound of the convex programs (\ref{cvxprogram}) and (\ref{cvxsymm})}

\begin{lemma}
\label{lemma2}
If $\{G^*,\vec{u}^*\}$ is an optimal feasible point of the optimization program (\ref{cvxprogram}), then the Bloch vectors $\{\vec n_x^*\}_x$ corresponding to $\vec{u}^*$ according to (\ref{udef}) are pure.
\end{lemma}

\noindent\textit{Proof.} --- Fix an index $i\in\{1,2,3\}$ and suppose $G^*_{ii}<1$. Define
\begin{equation*}
\widetilde G \;:=\; G^* + (1-G^*_{ii})E_{ii},    
\end{equation*}
where $E_{ii}$ has a $+1$ in the $(i,i)$ entry and zeros elsewhere. Then for any $\vec{x}\in\mathbb R^3$ we have
\begin{equation*}
\vec{x}^\top\widetilde G \vec{x} \;=\; \vec{x}^\top G^* \vec{x} + (1-G^*_{ii})x_i^2 \;\ge\; \vec{x}^\top G^* \vec{x} \ge 0,  
\end{equation*}
so $\widetilde G\succeq0$. The quadratic constraint for a variable $u^{pq,\pm}$ has the form
\begin{equation*}
(u^{pq,\pm})^2 \le G_{pp}+G_{qq}\pm 2G_{pq}. 
\end{equation*}

If the pair $(p,q)$ involves the index $i$, then the affine expression on the right-hand side is non-decreasing when $G^*$ is replaced by $\widetilde G$, because $G_{ii}$ appears with a coefficient $+1$ there. Therefore, every right-hand-side expression $G_{pp}+G_{qq}\pm 2G_{pq}$ is larger at $\widetilde G$ than at $G^*$ (due to $\widetilde{G}_{ii}=1>G^*_{ii}$). Constraints for pairs not involving $i$ remain unchanged.

Consequently, the feasible set of $u^{pq,\pm}$ variables corresponding to $\widetilde G$ contains the feasible set of those corresponding to $G^*$. Since the objective is linearly increasing in $u^{pq,\pm}$ and we are maximizing, enlarging the feasible set of $u^{ij,\pm}$ variables can only increase the optimal objective value. Therefore, there is only one optimal solution with the $i$-th diagonal equal to $+1$. Repeating this argument for each diagonal element yields an optimal Gram matrix with $G_{11}=G_{22}=G_{33}=1$. 
\qed

Based on the proof of Lemma~\ref{lemma2}, the convex program (\ref{cvxprogram}) can be solved by considering only pure states. In this case, the quadratic constraints take the following form:
\begin{equation*}
(u^{ij,\pm})^2 \leq 2(1 \pm G_{ij}), \quad ij \in \{12, 13, 23\}.    
\end{equation*}

Additionally, the constraint $0 \leq G_{ii} \leq 1$ should be replaced by $G_{ii} = 1$ for all $i \in \{1, 2, 3\}$, which reduces the number of constraints. Note that in~\eqref{cvxprogram} and~\eqref{cvxsymm}, we have not included any constraint on $G_{ij}$ for $i \neq j$. This omission is due to the fact that by specifying $G \succeq 0$ and $G_{ii} = 1$, the constraint $-1 \leq G_{ij} \leq 1$ is implicitly satisfied. 

\medskip

\noindent\textbf{Mirror-symmetric case.}---We state and prove the following lemma, which is similar to Lemma~\ref{lemma2}. 

\begin{lemma}
\label{lemma3}
If $\{G^*,\vec{u}^*\}$ is an optimal feasible point of the optimization program (\ref{cvxsymm}), then the Bloch vectors $\{\vec n_x^*\}_x$ corresponding to $\vec{u}^*$ according to (\ref{udef}) are pure.
\end{lemma}

\noindent\textit{Proof.} --- The proof of Lemma~\ref{lemma2} is applicable entirely except that $G_{ij}$ may differ from $G_{ik}$ in a resulting optimum. This would leave some of [C1] and [C2] unsaturated, thereby losing our ability to argue about the corresponding Bloch vectors. What follows is an indirect proof that $G_{ij} = G_{ik}$ in the optimum, i.e., $G_{ij}^* = G_{ik}^*$.

Let us assume that $G_{ij}^*=A<G_{ik}^*=B$; the opposite inequality can be treated similarly. Then, let us take two unit Bloch vectors $\vec n_i$ and $\vec n_j$ such that $G_{ij} = A = \vec n_i \cdot \vec n_j$. Next, choose a unit Bloch vector $\vec n_k$ such that $\vec n_j \cdot \vec n_k = G_{jk}^*$ and $\vec n_i \cdot \vec n_k = A$; such an $\vec n_k$ always exists in $\mathbb{R}^3$. Then, the Gram matrix of $\vec n_i$, $\vec n_j$ and $\vec n_k$ will be identical to $G^*$ except that $G_{ik}^*$ has been replaced by $G_{ik}=A=G_{ij}$. Taking this new Gram matrix as $G$ of Eq.~\eqref{cvxsymm}, the maximal value of $|u^{ij,+}|$ attainable according to [C1] and [C2] remains unchanged, but the value of $|u^{ij,-}|$ can be increased in the same way. Since the maximal $|u^{jk,\pm}|$ also remains unchanged, the value of $F_{ijk}$ can be increased by increasing $|u^{ij,-}|$ under the described change of $G$. This contradicts the assumption that the original $G^*$ was optimal. 
\qed

\subsection{The upper bound of $Q$ under a given mirror symmetry constraint}

We define $Q_{\text{mirror}}^{ijk}$ as the witness value $Q$ under a given symmetry constraint $(i,j,k)$. We consider the following functional, which is defined in terms of three unit Bloch vectors $\vec{n}_1,\vec{n}_2,\vec{n}_3 \in \mathbb{R}^3$:
\begin{equation}
\begin{aligned}
F &= \omega_{12}\lvert \vec{n}_1+\vec{n}_2\rvert + (1-\omega_{12})\lvert \vec{n}_1-\vec{n}_2\rvert \\
  &\quad+ \omega_{13}\lvert \vec{n}_1+\vec{n}_3\rvert + (1-\omega_{13})\lvert \vec{n}_1-\vec{n}_3\rvert \\
  &\quad+ \omega_{23}\lvert \vec{n}_2+\vec{n}_3\rvert + (1-\omega_{23})\lvert \vec{n}_2-\vec{n}_3\rvert.
\end{aligned}
\end{equation}

Since the mirror-symmetric bound is achievable by pure qubit states, the optimization can be restricted to pure states. In this case we may write
\begin{equation*}
\lvert \vec{n}_i \pm \vec{n}_j \rvert = \sqrt{2(1\pm c_{ij})},    
\end{equation*}
where $c_{ij} = \vec{n}_i \cdot \vec{n}_j$, which yields
\begin{align}
F(c_{12},c_{13},c_{23}) 
  &= \sqrt{2}\Big(\omega_{12}\sqrt{1+c_{12}} + (1-\omega_{12})\sqrt{1-c_{12}} \nonumber \\
  &\quad + \omega_{13}\sqrt{1+c_{13}} + (1-\omega_{13})\sqrt{1-c_{13}} \nonumber \\
  &\quad + \omega_{23}\sqrt{1+c_{23}} + (1-\omega_{23})\sqrt{1-c_{23}}\Big).
\end{align}

We now impose the symmetry constraint $c_{12}=c_{13}=c$ and denote $d := c_{23}$. The corresponding Gram matrix is
\begin{equation*}
 G = \begin{pmatrix}
1 & c & c \\
c & 1 & d \\
c & d & 1
\end{pmatrix}.   
\end{equation*}

\noindent Note that the condition $G \succeq 0$ is equivalent to the inequality
\begin{equation}\label{psd3}
d \ge 2c^2 - 1, \qquad c,d \in [-1,1],
\end{equation}
which guarantees nonnegative eigenvalues of $G$. Under these assumptions the functional takes the form
\begin{equation}\label{Fofcd}
F(c,d) = \sqrt 2 \,\big(f(c) + g(d)\big),    
\end{equation}
with
\begin{equation*}
f(c) = \alpha \sqrt{1+c} + \beta \sqrt{1-c}, 
\quad g(d) = p \sqrt{1+d} + q \sqrt{1-d},   
\end{equation*}
and coefficients
\begin{equation*}
 \alpha = \omega_{12}+\omega_{13}, \quad 
\beta = (1-\omega_{12})+(1-\omega_{13}), 
\end{equation*}
and
\begin{equation*}
p = \omega_{23}, \quad q = 1-\omega_{23}.   
\end{equation*}

Let us now consider two separate cases, depending on whether the maximum occurs at the boundary $d=2c^2-1$ in \eqref{psd3} or not. 

\subsubsection{\textbf{Case A}}

If the PSD constraint (\ref{psd3}) does not restrict the maximizers, i.e.\ if $c$ and $d$ can be chosen independently with $d \ge 2c^2-1$, the function $F$ decomposes additively. Then we may maximize $f$ and $g$ separately. By using the Cauchy--Schwarz inequality, we have
\[
f(c) \le \sqrt{2}\,\sqrt{\alpha^2+\beta^2}, 
\qquad 
g(d) \le \sqrt{2}\,\sqrt{p^2+q^2}.
\]
Hence
\begin{equation}\label{CSbound}
F(c,d) \le 2\sqrt{\alpha^2+\beta^2} + 2\sqrt{p^2+q^2} \;=:\; F^{123}_{\text{mirror}}.
\end{equation}
This bound is tight whenever
\begin{equation*}
c^* = \frac{\alpha^2-\beta^2}{\alpha^2+\beta^2}, 
\qquad 
d^* = \frac{p^2-q^2}{p^2+q^2}    
\end{equation*}
satisfy the feasibility condition $d^* \ge 2(c^*)^2 - 1$. In this case, the optimum coincides with the bound in Eq.~\eqref{CSbound}.

\subsubsection{\textbf{Case B}}

If $(c^*,d^*)$ obtained from independent maximization violates the PSD constraint (\ref{psd3}), then the maximum must occur on the boundary
\begin{equation*}
d = 2c^2 - 1,    
\end{equation*}
because $F$ has a single local maximum in the unconstrained $(c,d)$ plane according to (\ref{Fofcd}).

Substitution leads to the reduced one-variable optimization
\begin{align}
F(c)/\sqrt2 
&= \alpha\sqrt{1+c} + \beta\sqrt{1-c} 
   + 2p|c| + 2q\sqrt{1-c^2}.
\end{align}
Explicitly, this splits into
\begin{equation*}
 F_+(c)/\sqrt2 = \alpha\sqrt{1+c} + \beta\sqrt{1-c} + 2pc + 2q\sqrt{1-c^2}  
\end{equation*}
for $0 \le c \le 1$ and
\begin{equation*}
 F_-(c)/\sqrt2 = \alpha\sqrt{1+c} + \beta\sqrt{1-c} - 2pc + 2q\sqrt{1-c^2}
 \end{equation*}
for $-1 \le c \le 0$. The maximal value is then given by
\begin{equation*}
F^{123}_{\text{mirror}}=\max\Bigl\{\max_{c\in[0,1]} F_+(c), \;\; \max_{c\in[-1,0]} F_-(c)\Bigr\}.  
\end{equation*}
While closed-form expressions for the maximizers $c^*_\pm$ are in general algebraically cumbersome, they can be obtained efficiently by numerical methods, since $F_\pm(c)$ are smooth, one-dimensional functions. Moreover, $F_\pm(c)$ are strictly concave on the specified interval, and has at most one global maximizer. To obtain the mirror-symmetric bound, we note that $Q^{123}_{\text{mirror}} = \omega_{12}+\omega_{13} +\omega_{23} + F^{123}_{\text{mirror}}$.

The same analysis applies to the other symmetry constraints, namely, $c_{12}=c_{23}$ and $c_{13}=c_{23}$.

\section{Standard deviation of the witness value}\label{app1}

When measuring a qubit state, the probability of obtaining a particular outcome can be modeled by a binomial distribution. For a given number of trials (shots) $N$, the expected number of events corresponding to outcome $b\in\{0,1\}$ is
$N^{x}_{b|y} = N\, P(b|x,y)$,    
and the corresponding variance is
\begin{equation}
\sigma^{2}\bigl(N^{x}_{b|y}\bigr) = N\, P(b|x,y)\,(1-P(b|x,y)).    
\end{equation}

Since the witness is expressed in terms of expectation values, it is more convenient to consider the variance of the expectation value for an experiment defined by the preparation Bloch vector $\vec{n}_{x}$ and the measurement Bloch vector $\vec{m}_{y}$. In particular, the expectation value is given by $E_{xy} = P(0|x,y) - P(1|x,y)$,
and its variance is derived as follows:
\begin{equation}\label{varE}
\begin{aligned}
\sigma^{2}(E_{xy}) &= \sigma^{2}\Bigl(P(0|x,y) - P(1|x,y)\Bigr)\\[1mm]
&=\frac{1}{N^2}\,\sigma^{2}\Bigl(2N^{x}_{b=0|y} - N\Bigr)\\[1mm]
&=4\,P(0|x,y)\,(1-P(0|x,y))/N.
\end{aligned}
\end{equation}

Using Eq.~(\ref{varE}), we can now evaluate the variances of each witness $I_3(\omega_{ij})$. Specifically, for independent inputs $x$ and $y$ in the PM scenario, we have
\begin{align*}
\sigma^2\Bigl(I_3(\omega_{12})\Bigr) =&\, a_{12}\Bigl[\sigma^2(E_{11}) + \sigma^2(E_{21}) + \sigma^2(E_{41})\Bigr] \notag\\[1mm]
&+ b_{12}\Bigl[\sigma^2(E_{12}) + \sigma^2(E_{22}) \Bigr], \\
\sigma^2\Bigl(I_3(\omega_{13})\Bigr) =&\, a_{13}\Bigl[\sigma^2(E_{13}) + \sigma^2(E_{33}) + \sigma^2(E_{53})\Bigr] \notag\\[1mm]
&+ b_{13}\Bigl[\sigma^2(E_{14}) + \sigma^2(E_{34})\Bigr], \\
\sigma^2\Bigl(I_3(\omega_{23})\Bigr) =&\, a_{23}\Bigl[\sigma^2(E_{25}) + \sigma^2(E_{35}) + \sigma^2(E_{65})  \Bigr] \notag\\[1mm]
&+ b_{23}\Bigl[\sigma^2(E_{26}) + \sigma^2(E_{36})\Bigr],    
\end{align*}
where the coefficients are defined as $a_{ij}=\omega_{ij}^2$ and $b_{ij}=(1-\omega_{ij})^2$. Therefore, the variance of the overall witness $I_6$ is given by
\begin{align}
\sigma^{2}(I_6) =& \sum_{1\le i<j\le 3} \sigma^{2}\Bigl(I_3(\omega_{ij})\Bigr)
\nonumber\\
&=\frac{4}{N}\sum_{x,y} W_{xy}^2\, P(0|x,y)\,(1-P(0|x,y)),
\end{align}
and the associated standard deviation is
\begin{equation}
\Delta (I_6) = \sqrt{\sigma^{2}(I_6)}.
\end{equation}
\bibliographystyle{apsrev4-2} 
%

\end{document}